\documentclass[sigconf, nonacm]{acmart}
\AtBeginDocument{%
  }

\renewcommand\footnotetextcopyrightpermission[1]{} 
\begin{document}

\title{BERTDetect: A Neural Topic Modelling Approach for Android Malware Detection}

\author{Nishavi Ranaweera}
\orcid{1234-5678-9012}
\authornotemark[1]
\affiliation{%
  \institution{University of New South Wales}
  \city{Sydney}
  \country{Australia}
}

\author{Jiarui Xu}
\affiliation{%
  \institution{University of Sydney}
  \city{Sydney}
  \country{Australia}
}

\author{Suranga Seneviratne}
\affiliation{%
  \institution{University of Sydney}
  \city{Sydney}
  \country{Australia}
}

\author{Aruna Seneviratne}
\affiliation{%
  \institution{University of New South Wales}
  \city{Sydney}
  \country{Australia}
}

\renewcommand{\shortauthors}{Nishavi Ranaweera, Jiarui Xu, Suranga Seneviratne, and Aruna Seneviratne}
\begin{abstract}
Web access today occurs predominantly through mobile devices, with Android representing a significant share of the mobile device market. This widespread usage makes Android a prime target for malicious attacks. Despite efforts to combat malicious attacks through tools like Google Play Protect and antivirus software, new and evolved malware continues to infiltrate Android devices. Source code analysis is effective but limited, as attackers quickly abandon old malware for new variants to evade detection. Therefore, there is a need for alternative methods that complement source code analysis. Prior research investigated clustering applications based on their descriptions and identified outliers in these clusters by API usage as malware. However, these works often used traditional techniques such as Latent Dirichlet Allocation (LDA) and k-means clustering, that do not capture the nuanced semantic structures present in app descriptions. To this end, in this paper, we propose BERTDetect, which leverages the BERTopic neural topic modelling to effectively capture the latent topics in app descriptions. The resulting topic clusters are comparatively more coherent than previous methods and represent the app functionalities well. Our results demonstrate that BERTDetect outperforms other baselines, achieving $\sim$10\% relative improvement in F1 score.
\end{abstract}



\maketitle

\section{Introduction}\label{sec:Intro}

Mobile devices are pervasively used, and most of our online activities, be it web access, social networking, finance, gaming, or news reading, happen over mobile apps. While app market operators such as Google and Apple are working to make the mobile app ecosystem safer~\cite{Apple, Google}, mobile malware, spyware, and greyware are still being reported~\cite{news23, news24}. Most of the time, these apps with malicious intentions hide behind legitimate uses. For instance, according to recent reports, in 2023, Kaspersky identified over 153,000 malicious installation packages containing mobile banking Trojans~\cite{Kaspersky}. The rising prevalence of mobile malware also poses broader security challenges within corporate networks, as compromised smartphones and tablets can expose sensitive or proprietary data, potentially leading to data breaches.

Traditional methods of malware detection include creating signatures from source code patterns~\cite{malwareDetectionReview}. However, this method can not detect previously unseen malware and will have limitations when attackers create different variants or obfuscate code. Another approach is to monitor the run-time behaviour of apps, i.e., dynamic analysis and use anomaly detection methods~\cite{taintDroid}. This is a resource-intensive approach and can create many false positives~\cite{arp2014drebin}.

As a result, several works looked into the possibility of extracting signals from app metadata as a complementary means to detect new malware. For example, CHABADA~\cite{gorla2014CHABADA} clusters applications based on their natural language descriptions, and identify outliers by relying on the API usage within these clusters with the expectation of identifying suspicious or potentially malicious behaviour. It first uses Latent Dirichlet Allocation (LDA) to identify prevalent topics within the app descriptions. Based on the topic probabilities derived from LDA, it then applies the k-means clustering to cluster applications with similar descriptions.  Alecci et al.~\cite{alecci2024revisiting} follow a similar approach by using GPT embeddings of app descriptions. Both these studies treat the Google Play description of an app as a proxy for its advertised behaviour whilst using the API call sequences to verify the actual implemented behaviour. 

Recent advances in natural language processing, especially driven by transformer architectures, allow the possibility of improving these methods. Compared to traditional methods such as LDA, neural topic modelling has shown great success in many fields~\cite{wu2024, turan2024} in uncovering latent topics and themes from large amounts of text data. In this work, we show that neural topic modelling tools such as BERTopic~\cite{grootendorst2022bertopic} 
can be effectively used to discover common themes and underlying narratives in text data and as result, can be used to create a data analytics pipeline to detect Android malware. More specifically, we make the following contributions:
\begin{itemize}
    \item We propose BERTDetect, a novel framework for detecting Android malware using app metadata. BERTDetect utilizes BERTopic to cluster Google Play app descriptions into coherent topics. These clusters enable API pattern analysis to detect malicious behavior.  To our knowledge, this is the first application of neural topic modelling in malware detection.
    \item We compare BERTDetect's performance against baseline approaches such as LDA, CHABADA, and G-CATA, demonstrating significant improvements in malware detection. Notably, BERTDetect increases the True Positive Rate from 42.86\% to 50.89\% and the F1 score from 0.49 to 0.54 compared to these previous methods. 
    \item We conduct further analysis and attribute BERTDetect's effectiveness to more coherent topics generated by BERTopic. This coherence enables more precise outlier detection, allowing BERTDetect to accurately identify malicious apps that often advertises false pretences.
\end{itemize}

\section{Related Work}\label{sec:related}

\begin{figure*}[ht!]
\centering
      \includegraphics[width=\linewidth]
      {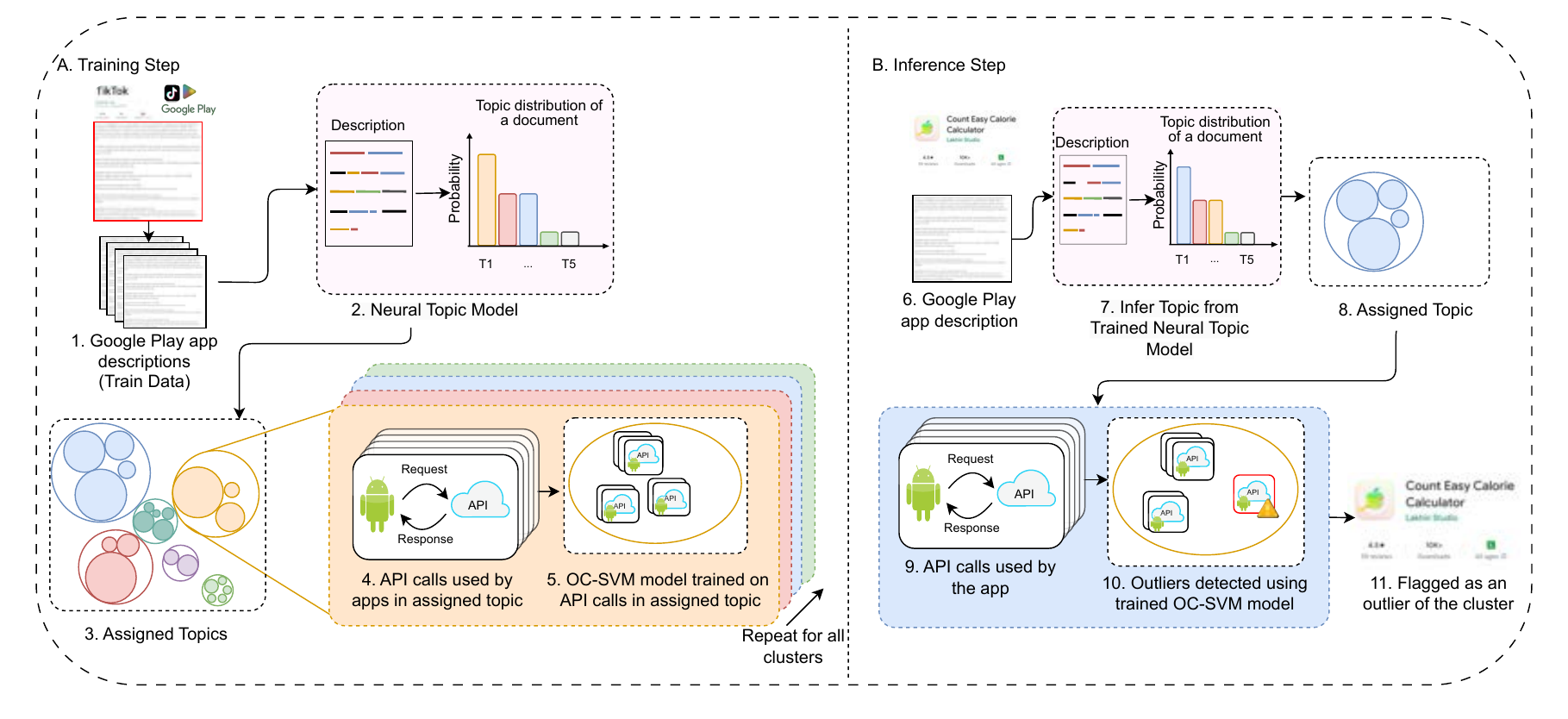}
      \vspace{-2mm}
\caption{BERTDetect Framework} 
\Description{}
\vspace{-4mm}
\label{fig:method}
\end{figure*}

\subsection{Android Malware Detection}
Many studies have shown that mobile apps often request permissions that are unrelated to their intended functionality. This behavior not only poses significant privacy risks but also serves as an indicator for detecting malicious activities \cite{book2013longitudinalanalysisandroidad, permissions3, demistyfiedAndroid, permissions2, permissions1}.
DREBIN~\cite{arp2014drebin} is one of the early work in Android malware detection that allows lightweight smartphone analysis. It combines static analysis of app features—permissions, API calls, and network addresses—into a vector space and uses machine learning to classify apps as malicious or benign. Other static analysis approaches such as~\cite{droidmat2012,mariconti2016mamadroid} have also contributed significantly to the development of malware detection techniques. In order to overcome the limitations of static methods and improve detection accuracy, dynamic analysis approaches, such as DroidScope~\cite{droidscope2012} and  TaintDroid~\cite{taintDroid} use dynamic taint analysis, and Andromaly~\cite{shabtai2012andromaly} uses anomaly detection techniques to detect malware at runtime. 

While dynamic approaches can detect malware that evades static analysis, they are often resource-intensive, and require a controlled execution environment. Other approaches address these challenges by integrating both static and dynamic features with machine learning models to improve detection performance~\cite{droidsec2014,alzaylaee2020dldroid}.
For instance, DL-Droid~\cite{alzaylaee2020dldroid}, is a deep learning-based system that improves Android malware detection by carefully generating test inputs based on the app's current behavior. DL-Droid demonstrated high detection rates, particularly in identifying new, unseen malware, achieving up to 99.6\% accuracy when using both dynamic and static features. However, this approach, while effective, adds complexity and increases computational demands, which may limit its practicality in environments with limited resources.

\textit{In summary, mainstream Android malware detection methods—static and dynamic analysis—each have limitations. Static analysis struggles with code changes and obfuscation, while dynamic analysis demands significant computational resources. These challenges underscore the need for complementary strategies, like using metadata, to improve detection.}

\subsection{Metadata for Android Malware Detection}

To overcome the limitations of source code-based methods and dynamic analysis, other works have explored the inclusion of easily accessible metadata. For example, CHABADA~\cite{gorla2014CHABADA} leverages the disparity between the advertised and actual behavior of the app to detect anomalies. It uses Latent Dirichlet Allocation (LDA) for topic modelling and subsequent clustering of Android app descriptions, coupled with an analysis of sensitive API usage via unsupervised One-Class SVM. Due to its unsupervised nature, CHABADA can detect malicious behaviours without prior knowledge of malware signatures.

Recent work G-CATA~\cite{alecci2024revisiting} generates OpenAI text embeddings of app descriptions to directly cluster these embedding using k-means algorithm. By doing so, G-CATA achieves more accurate grouping of apps that lead to a significant improvement in malware detection performance compared to CHABADA. However, it may not capture topics as effectively as a specialised topic modelling approach, which utilize iterative processes to optimize topic assignments. A few other works used app metadata to detect non-malicious yet dubious behaviors of mobile apps. These include detecting spam apps~\cite{seneviratne2015early-detection, seneviratne2017spam}, deliberate mis-categorizations~\cite{surian2017app-miscategorization}, and app counterfeits~\cite{karunanayake2020multi,rajasegaran2019multi}.

\textit{In contrast to these, BERTDetect goes a step further by integrating neural topic modelling to cluster apps based on their app descriptions. Rather than relying solely on static metadata or predefined categories, our framework effectively identifies outliers within these clusters by analyzing API call patterns, offering a more effective and adaptable method for mobile malware detection.}

\section{BERTDetect Framework}
\label{sec:Methodology}
BERTDetect is based on the intuition that an app's functionality can be represented by one or more high-level natural language topics. For example, a food delivery app may consist of functionality groups related to food defined by words such as \texttt{\{food, order, dinner, pizza\}} and maps and navigation defined by words such as \texttt{\{location, map, co-ordinates\}} for the food delivery part. These functionality groups can be associated with various Android APIs~\cite{gorla2014CHABADA}. For example, the delivery-related functionality group will use Android's \texttt{Location Manager} API calls. And the online ordering functionality group will access \texttt{Subscriptions and In-App Purchases} API. If an app accesses API calls outside its functionality groups, it can be considered an outlier and potentially malicious.

Based on this intuition, we propose BERTDetect, Android malware detection framework, which has two main phases: \textit{Training} and \textit{Inference}. The training phase consists of three steps; {\textit{i) Generating topics for Google Play descriptions using BERTopic}}, {\textit{ii) Assigning apps to topics based on topic affinity}}, and {\textit{ iii) Training One-Class SVMs for each topic based on their API call usage.}} In the inference phase, new apps are evaluated against the established topic clusters and the corresponding outlier detection model to identify potential malware.  In Figure~\ref{fig:method}, we show a schematic overview of our process. Next, we describe each step of the training phase in detail, followed by the inference phase.

\subsection{Training Phase}

\subsubsection{Generating topics for Google Play descriptions using BERTopic}\label{SubSubSec:GenTopic}

We first generate a set of topics and a topic probability distribution by training a BERTopic model~\cite{grootendorst2022bertopic} using Google Play descriptions of benign apps. A topic probability distribution indicates the likelihood that a particular app description is associated with each of the topics, providing a way to quantify the relevance of topics within the text.

BERTopic first creates document embeddings for app descriptions using a pre-trained BERT \cite{devlin2018bert} model  to obtain document-level information. 
Then, the dimensionality of these embeddings is reduced using the UMAP \cite{mcinnes2018umap} algorithm, and similar embeddings are clustered together using the HDBSCAN algorithm. Finally, the app descriptions of each cluster are tokenized and weighed using class-based variation of TF-IDF to generate a topic representation.

We chose BERTopic for this module of the framework because it effectively handles the non-spherical nature of clusters, providing more accurate and meaningful topic representations. This capability is especially useful for capturing the complex structure of textual data, as evidenced by BERTopic's superior clustering performance in domains such as news classification and scientific literature analysis~\cite{bertopicNews,wang2023identifying}. Other topic modelling methods such as Top2Vec \cite{angelov2020top2vec} and Doc2Vec \cite{le2014ddoc2vec}, use a centroid-based topic modelling approach that generates the topic representation from words that are close to the cluster's centroid. This centroid-based approach assumes that clusters are spherical, centered around a single point. However, in real-world scenarios, clusters often have more complex, non-spherical structures, and relying on centroids can lead to less accurate and potentially misleading topic representations. BERTopic's ability to capture these complex cluster shapes makes it a good fit for our framework.

BERTopic automatically determines a fixed number of topics for any given dataset, which is particularly advantageous for applications where the exact number of topics is not known beforehand. This number is determined based on the clustering method employed by BERTopic and the specific characteristics of the dataset being analyzed. When we apply BERTopic to our dataset of benign app descriptions, it identifies 76 distinct topics, a number that is comparable to those used in other works, such as G-CATA, which also utilize a similar range of topics ({\bf cf.} Section~\ref{SubSec:Baseline}).

\subsubsection{Assigning apps to topics based on topic affinity}\label{open}

\begin{figure}[htbp]
\centering
      \includegraphics[width=\linewidth]
      {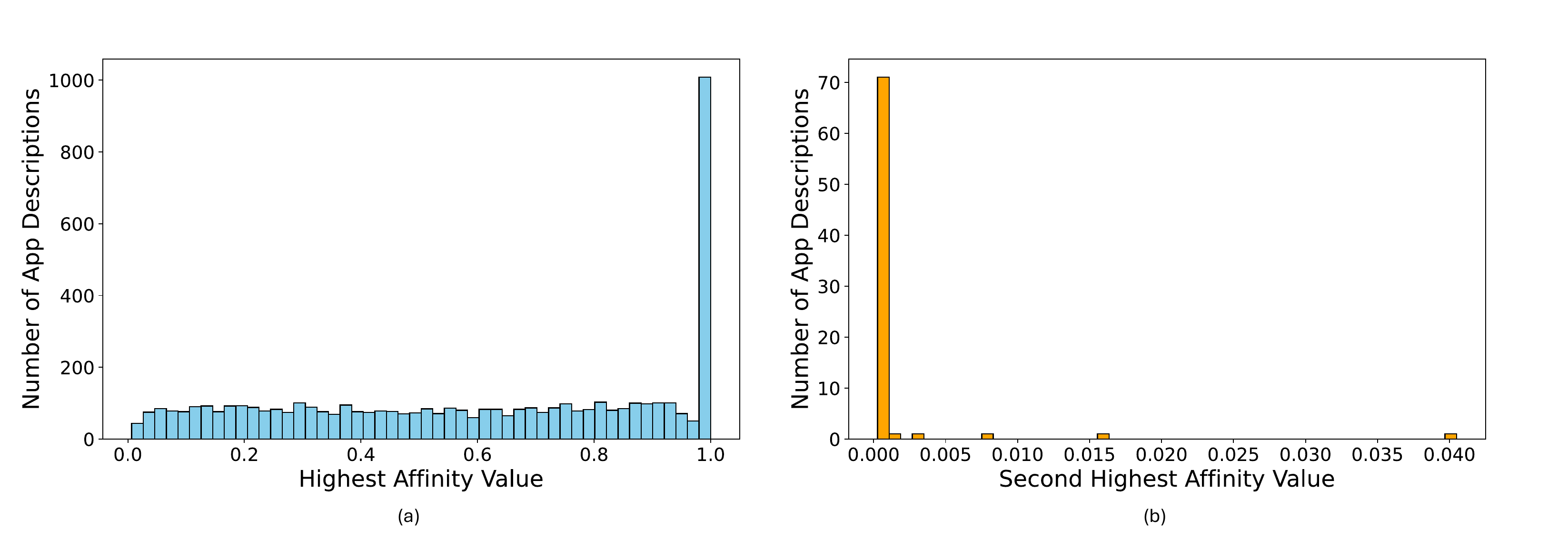}
      \vspace{-2mm}
\caption{Distribution of the first and second highest affinity values assigned to app descriptions in BERTopic. (a) illustrates that, for most app descriptions, the highest affinity values are close to 1.0, indicating a strong association with a single dominant topic. (b) displays the distribution of the second highest affinity values, which are predominantly close to zero.} 
\Description{}
\vspace{-4mm}
\label{fig:topic_affinity}
\end{figure}

Topic modelling associates an app description with each topic, assigning a specific probability to indicate the strength of that association. Similar to the usage in CHABADA~\cite{gorla2014CHABADA}, we refer to this as ``topic affinity". Our next step is to assign each app description, and thus the corresponding app, to a single topic by selecting the topic with the highest probability, i.e., the topic with the strongest affinity. As shown in Figure~\ref{fig:topic_affinity}, BERTopic tends to allocate a high probability to a single dominant topic for most app descriptions. This suggests that the model is confident in its assignment, making it reasonable to rely on the topic with the highest affinity as the primary representation of the app. This clear assignment ensures that each app is associated with the topic that best captures its core characteristics. It provides an interpretable representation that minimizes ambiguity. By grouping apps based on their dominant topic, we can identify groups of applications showing similar characteristics, enabling us to analyze patterns of behavior and usage within these topics. We present qualitative examples in Appendix~\ref{appendix:topic_wordclouds}, comparing the topics generated for Viber and Telegram across BERTopic, LDA, and CHABADA. These examples highlight BERTopic’s effectiveness in capturing coherent and interpretable topics that align with each app’s primary functionality. 

\subsubsection{Training One-Class SVMs for each topic based on their API call usage}\label{var_len} 
For each app assigned to a given topic, we extract the API calls by analyzing the app's APK file using Androguard~\cite{androguard}. We then use the same method as Alecci et al.~\cite{alecci2024revisiting} to extract the list of used API calls that are protected by permissions. Then, we convert these API calls into binary feature vectors to train One-Class Support Vector Machine models (OC-SVM) for each topic.  As mentioned in Section~\ref{SubSubSec:GenTopic}, during training, we use only benign apps. 

\subsection{Inference Phase}
As illustrated in Figure~\ref{fig:method}(B), during the inference phase, we use the trained BERTopic model to find the topic for each app in the test set. Then, for each app, we extract the API calls and use the corresponding OC-SVM to decide whether the app belongs to that class or not - if not, we consider it malware. Unlike the training phase, which used only benign apps, here we use benign and known malicious apps to assess the performance of BERTDetect and compare with baselines.
\section{Experiment Setup and Results}
\label{sec:Results}

In this section, we introduce the dataset used to evaluate BERTDetect's performance. Next, we describe the baselines used for comparison, followed by malware detection results.

\subsection{Dataset}
We use the \textsc{AndroCatSet} dataset by Alecci et al.\cite{alecci2024revisiting}, comprising 5,000 benign apps from the Google Play Store, categorized into 50 unique classes (100 apps per class), manually verified by the authors. Each app includes a unique app ID, description text, and a list of sensitive API calls, governed by Android permissions and extracted via Androguard\cite{androguard}.
 
\textsc{AndroCatSet} also includes 500 malicious apps. However they did not contain the app descriptions, which we extracted from AndroZoo~\cite{androzoo} and the Google Play Store. Out of the 500 malicious samples, descriptions for 448 were successfully retrieved.

We split \textsc{AndroCatSet} into training, validation, and test sets (Table~\ref{Table:DatasetDistribution}). The training set contains 4,000 benign apps, while validation and test sets each have 500 benign apps, expanded with 224 malicious apps randomly sampled. The training set contains only benign apps, since our unsupervised framework learns from them exclusively.

\begin{table}[H]
\scriptsize
\centering
\caption{Dataset Summary} 
\vspace{-4mm}

\begin{tabular}{p{2.5cm}p{1.5cm}p{1.5cm}p{1.5cm}}\specialrule{.12em}{1em}{0em}

{\bf{Dataset Split}} & {\bf Benign Apps} & {\bf Malicious Apps} & {\bf Total Apps} \\ 
\hline

Training Set  & 4000 & - & 4000 \\ 

Validation Set  & 500 & 224 & 724 \\ 

Test Set  & 500 & 224 & 724 \\
\specialrule{.12em}{0em}{0em}
\end{tabular} 
\label{Table:DatasetDistribution}
\end{table}

\subsection{Baseline Models}
\label{SubSec:Baseline}
We compare our method with three baselines.

\noindent{{\bf i) LDA~\cite{lda2003}} ({\bf L}atent {\bf D}irichlet {\bf A}llocation) A probabilistic topic model that identifies latent topics in text. Using the LDA MALLET library,\footnote{LDA MALLET (MAchine Learning for LanguagE Toolkit)} apps are assigned to the topic with the highest affinity.}

\noindent{{\bf ii) CHABADA~\cite{gorla2014CHABADA}}, Combines LDA with k-means clustering on topic affinities of app descriptions. It identifies outliers using OC-SVMs trained on sensitive APIs. For fair comparison, we set the number of topics to 50, following Alecci et al.~\cite{alecci2024revisiting}.

\noindent{{\bf iii) G-CATA~\cite{alecci2024revisiting}} ({\bf G}PT-based {\bf CAT}egorization of {\bf A}ndroid apps) 
Generates app description embeddings using OpenAI’s \textit{ada text embeddings API} and clusters them into 50 groups via k-means. OC-SVMs trained on benign app APIs are used to detect outliers. Unlike LDA, CHABADA, and BERDetect, G-CATA clusters directly from text embeddings.}

Finally, we highlight that unlike CHABADA and G-CATA, BERTDectect which uses BERTopic, does not involve a separate clustering step. This is because BERTopic integrates clustering within its topic modelling process ({\bf cf.} Section~\ref{SubSubSec:GenTopic}).

\subsection{Malware Detection Performance}

We present the results in Table~\ref{Table:OCSVM}, where BERTDetect emerges as the most effective method, achieving the highest F1 Score of 0.54 and a TP Rate of 50.89\%, indicating its ability to identify malware while maintaining a balanced performance across other metrics. 
\begin{table}[H]
\scriptsize
\centering
\caption{Performance Results for Malware Detection} 
\vspace{-4mm}
\begin{tabular}{p{1.3cm}p{0.9cm}p{0.9cm}p{0.9cm}p{0.9cm}p{0.9cm}p{1.9cm}}\specialrule{.12em}{1em}{0em}

{\bf{Method}} & {\bf TN Rate} & {\bf FP Rate} & {\bf FN Rate}& {\bf TP Rate}& {\bf F1 Score}\\ 
\hline

LDA  & 85.60\%
 & 14.40\% & 80.80\% & 19.20\% & 0.25
\\ 

CHABADA  & 83.80\%
 & 16.20\% & 68.75\% & 31.25\% & 0.37
\\ 

G-CATA   & {\bf 86.60}\% 
 & {\bf 13.40\%} & 57.14\% & 42.86\% &0.49
\\ 

BERTDetect  & 82.40\%
& 17.60\% & {\bf 49.11\%} & {\bf 50.89\%} &{\bf 0.54}
\\

\specialrule{.12em}{0em}{0em}
\end{tabular} 
\label{Table:OCSVM}
\end{table}

Although G-CATA exhibits the highest True Negative (TN) rate at 86.60\%, it shows low True Positive (TP) rate of 42.86\%. In malware detection, it is crucial to maintain a high true positive rate while also achieving a high true negative rate to avoid missing real threats and minimizing false alarms. The significant increase of the True Positive Rate in BERTDetect compared to G-CATA, and the overall improvement of F1 score from 0.49 to 0.54 (i.e, $\sim$10\% relative increase), confirms the ability of BERTopic to more accurately differentiate between benign and malicious apps. 

\section{Results Analysis}
\label{sec:Analysis}

We conduct analysis on the topic and cluster quality of BERTDetect and other baselines to explain why BERTDetect is performing better.

\subsection{Topic Quality Evaluation}
\label{subsection:topic_quality}

First, we compare the quality of topics generated by BERTDetect and other baselines using topic coherence measures. \textit{Topic coherence} reflects the human-interpretability of topics, i.e., the degree to which the words within a topic cluster are semantically related~\cite{lau-etal-2014-machine,newman2010}. A high topic coherence indicates that the topics generated are meaningful and can be easily understood, which ensures that they (or clusters formed from them) accurately represent distinct and interpretable themes of app functionalities. Comparable to other work~\cite{grootendorst2022bertopic,roder2015}, we measure the topic coherence using \textit{NPMI} and \textit{Cv}. We provide a brief description of each metric in Appendix \ref{appendix:metrics}. 

Figure~\ref{fig:cv} presents complementary cumulative distribution function (CCDF) plots of \textit{NPMI} and \textit{Cv} scores for BERTopic, LDA, and CHABADA. Note that we don't have these results for the G-CATA because it applies k-means clustering directly to text embeddings. As a result, it does not generate topics that can be evaluated using coherence measures like \textit{NPMI} and \textit{Cv}.

According to Figure~\ref{fig:cv}(a), BERTopic consistently achieves higher \textit{NPMI} scores compared to LDA and CHABADA, indicating superior topic quality. Around 70\% of the topics generated by BERTopic have a \textit{NPMI} score greater than 0.2, whereas around 40\% of the topics from LDA and 0\% from CHABADA achieve \textit{NPMI} scores higher than 0.2. This difference illustrates BERTopic’s ability to produce more coherent topic clusters as measured by the \textit{NPMI} metric.

Similarly, in Figure~\ref{fig:cv}(b), which presents the CCDF of \textit{Cv} coherence scores, BERTopic also outperforms LDA and CHABADA. Approximately 90\% of the topics generated by BERTopic have a \textit{Cv} score above 0.6, while only about 60\% of LDA’s topics and none from CHABADA exceed this threshold. This further demonstrates the robustness of BERTopic in generating semantically coherent topics across different coherence measures.
Overall, the CCDFs of BERTopic demonstrate a slower decline, implying that a higher proportion of its topics attain higher coherence scores. In contrast, LDA and CHABADA show steeper declines, with CHABADA exhibiting the lowest overall topic quality. 

\begin{figure}[ht!]
\centering
      \includegraphics[width=\columnwidth]
      {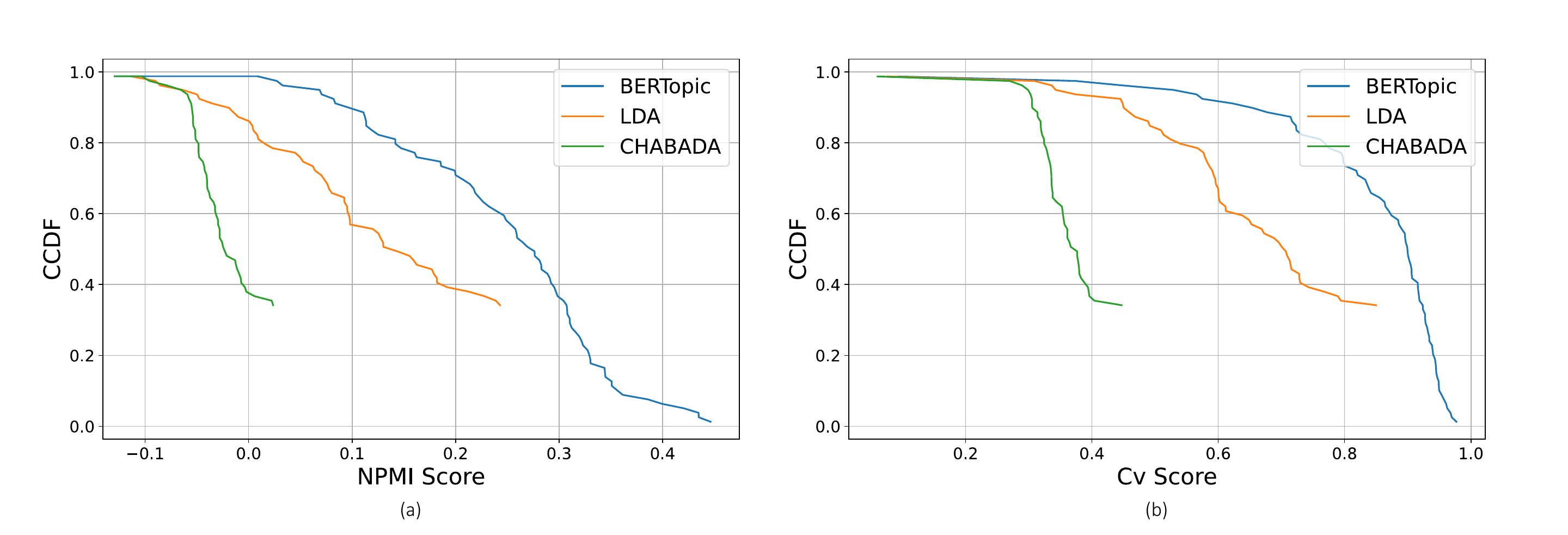}
      \vspace{-2mm}
\caption{CCDF of Topic Cluster Quality. The curves demonstrate how BERTopic consistently  maintains higher quality topic clusters across different metrics compared to the baselines LDA and CHABADA.} 
\Description{}
\vspace{-4mm}
\label{fig:cv}
\end{figure}

Despite showing poor topic quality, earlier, we saw CHABADA performed better in the end-to-end malware detection task than LDA ({\bf cf.} Table~\ref{Table:OCSVM}). This is because CHABADA performs a clustering step after applying LDA for topic modelling, optimizing specifically for the objective of distinguishing outliers rather than generating semantically coherent topics. This additional clustering step allows CHABADA to cluster different topics together, capturing similar app behaviors necessary for outlier detection.
\textit{Overall, the results show that BERTopic consistently outperforms both LDA and CHABADA in terms of topic coherence, as evidenced by higher average scores and relatively low standard deviations for both \textit{Cv} and \textit{NPMI} metrics. LDA, while performing better than CHABADA, shows a notable gap in coherence quality compared to BERTopic. As we further demonstrate later, the higher topic quality of BERTopic is the main contributing factor to the higher malware detection performance of BERTDetect.}

\subsection{Effect of Topic Quality}

\begin{figure}[ht!]
\centering
      \includegraphics[width=\columnwidth]
      {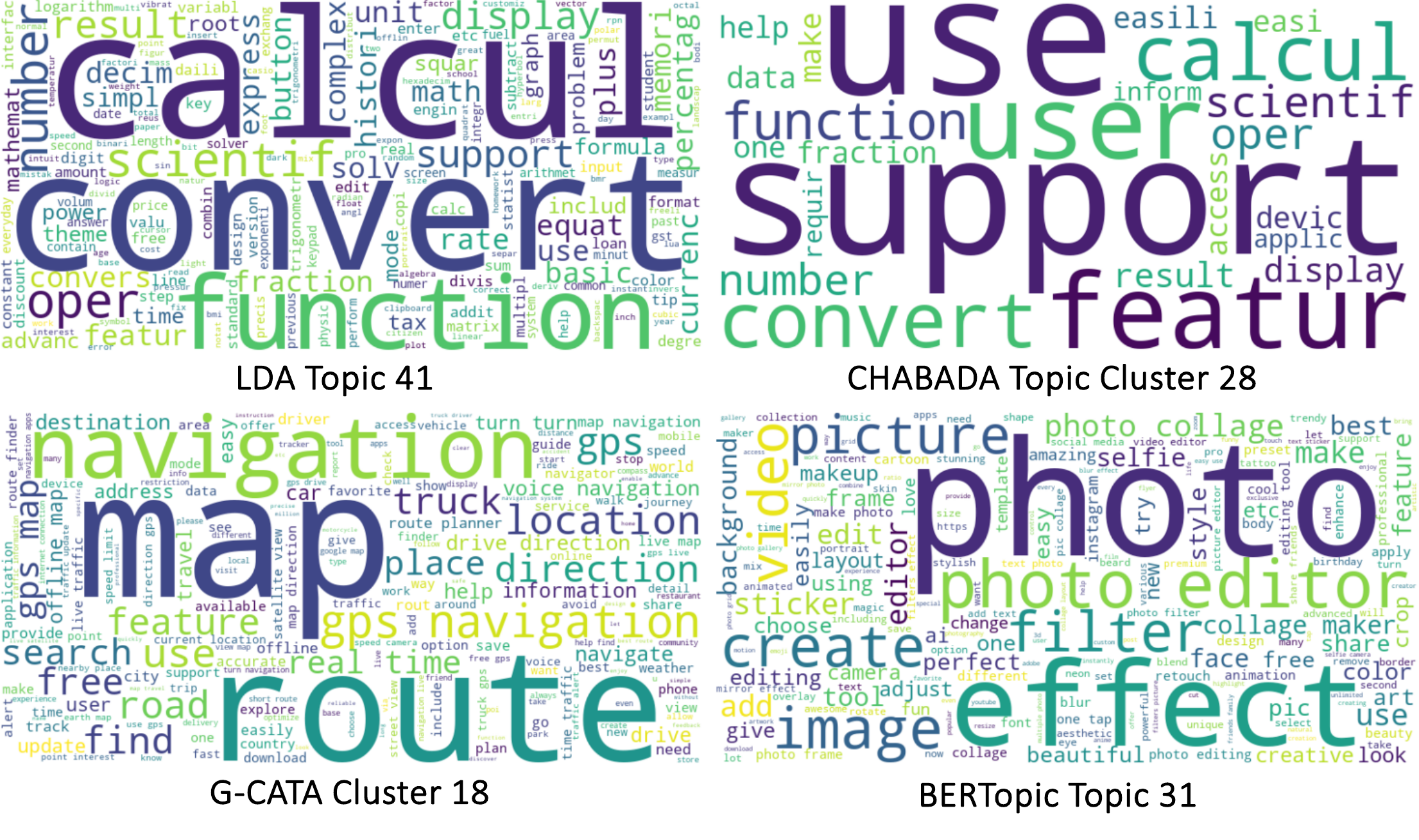}
      \vspace{-2mm}
\caption{Word clouds of the topic assignments of a malicious ``Bridal makeup" app.} 
\Description{}
\vspace{-4mm}
\label{fig:bridal_makeup_wordcloud}
\end{figure}

\begin{figure}[ht!]
\centering
      \includegraphics[width=\columnwidth]
      {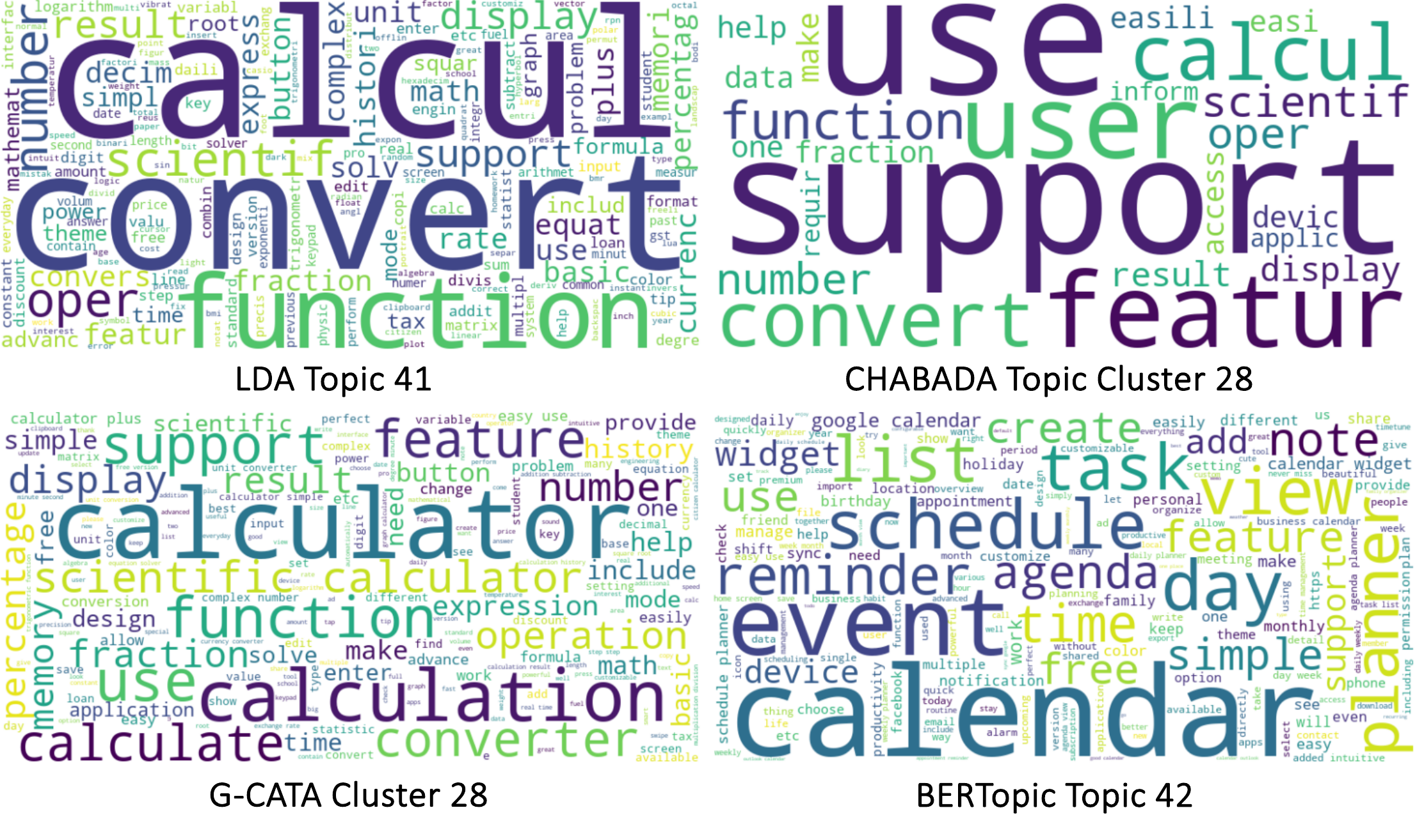}
      \vspace{-2mm}
\caption{Word clouds of the topic assignments of a malicious ``Event reminder" app.} 
\Description{}
\vspace{-4mm}
\label{fig:event_calendar_wordcloud}
\end{figure}

Figures~\ref{fig:bridal_makeup_wordcloud} and~\ref{fig:event_calendar_wordcloud} show two malware topic assignment examples. They illustrate the word clouds of the topics to which two malware apps were assigned, based on their app description by LDA, CHABADA, G-CATA, and BERTDetect. 

For instance, as shown in Figure~\ref{fig:bridal_makeup_wordcloud}, a malicious app with a description related to bridal makeup is assigned by BERTopic to the topic ``photo, editor, effects,  makeup, and filters", which aligns well with the app's description. However, CHABADA and LDA categorize this app under clusters related to mathematical operations and conversions, while G-CATA places it in a cluster associated with navigation. Similarly, in Figure~\ref{fig:event_calendar_wordcloud}, an event reminder and tracker app is correctly assigned by BERTopic to the topic about ``calendar, agenda, events, and reminders," while the other models incorrectly classify it under unrelated clusters like calculator or conversion tools. These examples illustrate the limitations of CHABADA, LDA, and G-CATA in accurately capturing the semantic meaning of app descriptions, whereas BERTopics's assignments show better contextual alignment, suggesting it as a more reliable method for identifying malware at the OC-SVM step.

\begin{figure}[ht!]
\centering
      \includegraphics[width=\linewidth]
      {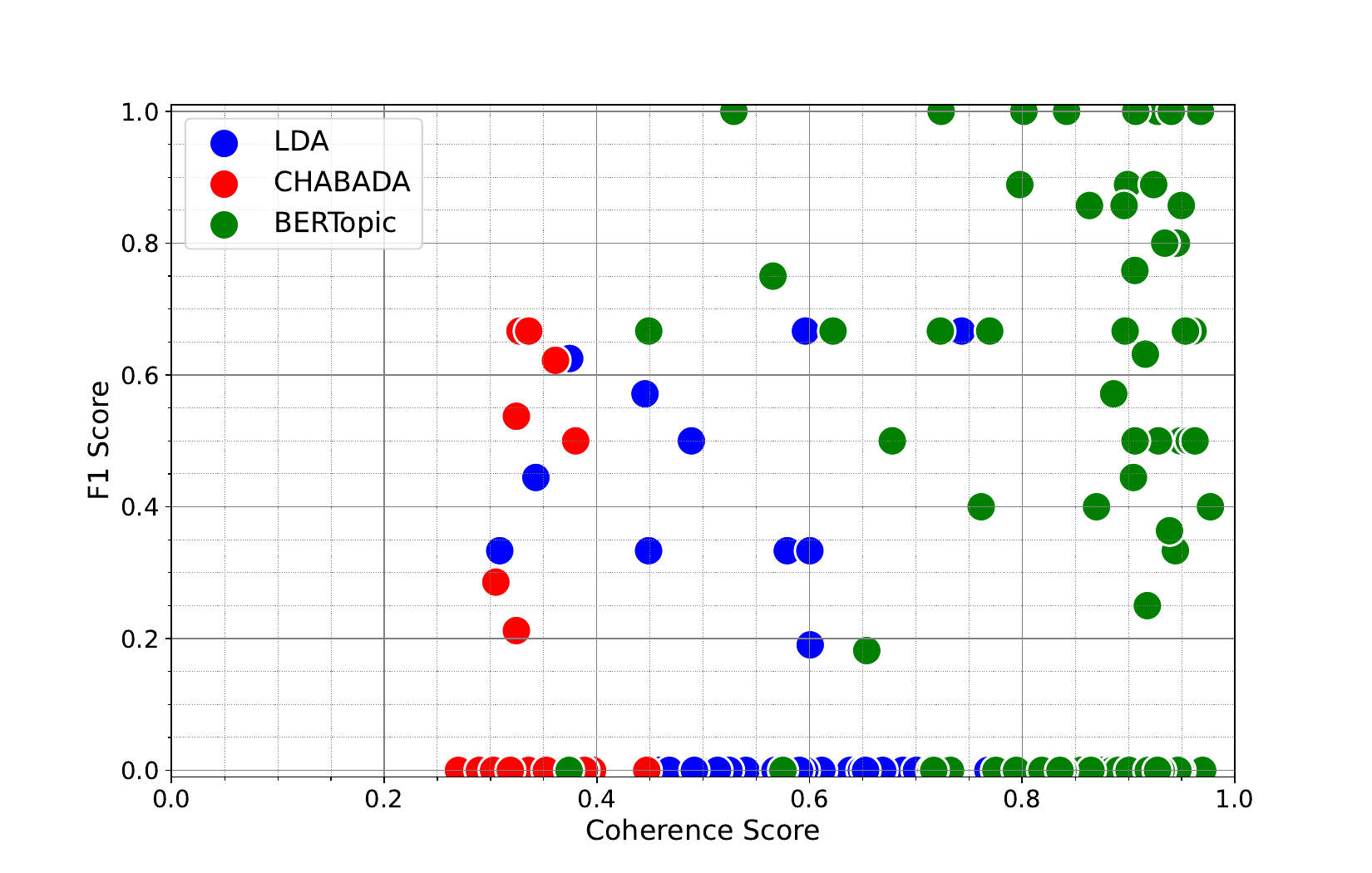}
      \vspace{-2mm}
\caption{Cv Coherence vs F1 Score of individual topic clusters from topic models LDA, CHABADA and BERTopic. } 
\Description{}
\label{fig:scatter}
\end{figure}

Based on this observation, we next check whether the coherence of the topic/topic cluster has an impact on the performance of the malware detection results of the OC-SVMs. The scatter plot in Figure~\ref{fig:scatter} shows the cluster/topic-wise F1 scores for LDA,  CHABADA, and BERTopic. We can see how both LDA and CHABADA have most of their clusters with very low F1 scores, and only a limited number of clusters out of the 50 clusters have non-zero F1 scores. In contrast, BERTopic has more topics with higher F1 scores, and they are mostly aligned to the right side of the x-axis, indicating higher coherence. Hence, the topics generated by BERTopic show an overall higher coherence and F1 score in a majority of its topics.

All in all, these findings highlight the significant advantage of BERTopic over both LDA and CHABADA in generating topic clusters that are not only more coherent but also lead to better outlier detection performance when applied with OC-SVMs. The higher alignment of BERTopic's clusters on the right side of the x-axis in the scatter plots indicates a strong correlation between topic coherence and F1 scores. This suggests that BERTopic's ability to produce more meaningful and distinguishable topics is a contributing factor to its superior performance in identifying outliers. 

\subsection{Ablation w.r.t Neural Topic Modelling} 
Finally, to show that neural topic modelling of app descriptions has a tangible impact on outlier detection performance, we trained an OC-SVM model using API calls extracted solely from benign apps. This approach allows to establish a baseline for understanding how well the classifier performs without neural topic modelling, thereby isolating the effects of topic modelling on the detection process. We present the performance of this model on our test set in Table~\ref{Table:ablation}.

\begin{table}[H]
\scriptsize
\centering
\caption{Results with and without Neural Topic Modelling} 
\vspace{-4mm}
\begin{tabular}{p{1.6cm}p{0.9cm}p{0.9cm}p{0.9cm}p{0.9cm}p{0.9cm}p{1.9cm}}\specialrule{.12em}{1em}{0em}

{\bf{Method}} & {\bf TN Rate} & {\bf FP Rate} & {\bf FN Rate}& {\bf TP Rate}& {\bf F1 Score}\\ 
\hline

OC-SVM only  & 55.60\% & 44.40\% & 29.46\% & 70.54\% & 0.52 \\ 

BERTDetect  & 82.40\%
& 17.60\% & {49.11\%} & { 50.89\%} &{\bf 0.54}
\\ 

\specialrule{.12em}{0em}{0em}
\end{tabular} 
\label{Table:ablation}
\end{table}

The results from the OC-SVM model trained solely on benign API calls demonstrate notable limitations in distinguishing between malicious and benign applications. While the model achieves a relatively high True Positive Rate (TP Rate) of 71\%, its True Negative Rate (TN Rate) is considerably low at 56\%, indicating that nearly half of the benign apps are incorrectly classified as malicious. This is further reflected in the high False Positive Rate (FP Rate) of 44\%, which suggests that many benign apps are mistakenly flagged as threats. In inherently imbalanced applications, such as malware detection, where the consequences of false positives are high (e.g., generating unnecessary alerts disrupting user experience or creating alert fatigue), F1 score, which ignores True Negatives, may not comprehensively evaluate the model's effectiveness. As a result, even though the F1 scores appear close, BERTopic outperforms the basic model without topic modelling as it provides a better balance between the true positives and true negatives. Finally, we also highlight that when compared to other baselines in Table~\ref{Table:OCSVM}, the naive OC-SVM method, in fact, performs well, surpassing all of them, again highlighting the inadequacy of topic modelling approaches of those works.
\section{Discussion and Concluding Remarks}
\label{sec:Conclusion}

In this paper, we proposed BERTDetect, a neural topic modelling approach that can identify functionality clusters within Google Play descriptions of Android apps. These clusters enhanced the accuracy and robustness of Android malware detection, providing a balance between true positives and true negatives. As demonstrated in Section~\ref{sec:Results}, our approach using BERTopic consistently outperformed baselines such as LDA, CHABADA, G-CATA. More specifically, BERTDectect achieved a 10\% relative increase in F1 score compared to the next closest performing model G-CATA. We also provided quantitative and qualitative evidence that the higher performance of our method can be attributed to BERTopic's (underlying topic model behind BERTDetect) ability to generate more coherent and meaningful topics, representing app functionalities correctly. Overall, our approach expands upon conventional methods, delivering more reliable detection of malicious apps. We next discuss the limitations and possible future improvements of our work. \\

\noindent{{\bf Use of app metadata for malware detection:} BERTDetect complements existing static and dynamic analysis methods by leveraging app metadata, offering a lightweight approach that avoids parsing source code or monitoring network activities. While stronger signals can be obtained from code-level or runtime analysis, BERTDetect can enhance overall protection—especially against previously unknown malware—when combined with these methods.}

\noindent{{\bf Impact of neural topic modelling:} 

We used BERTopic as our primary model, but other neural topic models can provide similar benefits for outlier detection. Our analysis in Section IV(F) confirms that topic modeling is crucial for boosting precision and recall. Notably, BERTopic outperformed the Large Language Model (LLM)-based GPT embeddings, and future work could explore other neural topic models such as CombinedTM~\cite{bianchi-etal-2021-pre}, TopClus~\cite{topclus2022}, or FASTopic~\cite{wu2024fastopicfastadaptivestable}}

\noindent{{\bf Limitations of the dataset:} Comparable to prior work~\cite{alecci2024revisiting, gorla2014CHABADA}, we demonstrated the performance of BERTDetect on the \textsc{AndroCatSet} dataset.
However, there is a notable lack of datasets containing both APK files and corresponding app descriptions. Other Android malware datasets~\cite{arp2014drebin,maldroid2020} only include APK files, making it difficult to test metadata-based methods like BERTDetect on a broader scale.}

\section{Acknowledgments}
This research was supported by the Australian Government through the Australian Research Council's Discovery Projects funding scheme (Project ID DP220102520).


\bibliographystyle{plainnat}
\bibliography{biblio}

\begin{thebibliography}{46}
\providecommand{\natexlab}[1]{#1}
\providecommand{\url}[1]{\texttt{#1}}
\expandafter\ifx\csname urlstyle\endcsname\relax
  \providecommand{\doi}[1]{doi: #1}\else
  \providecommand{\doi}{doi: \begingroup \urlstyle{rm}\Url}\fi

\bibitem[Alecci et~al.(2024)Alecci, Samhi, Bissyand{\'e}, and Klein]{alecci2024revisiting}
Marco Alecci, Jordan Samhi, Tegawend{\'e}~F Bissyand{\'e}, and Jacques Klein.
\newblock {Revisiting Android App Categorization}.
\newblock In \emph{Proceedings of the IEEE/ACM 46th International Conference on Software Engineering}, pages 1--12, 2024.

\bibitem[Allix et~al.(2016)Allix, Bissyand{\'e}, Klein, and Le~Traon]{androzoo}
Kevin Allix, Tegawend{\'e}~F. Bissyand{\'e}, Jacques Klein, and Yves Le~Traon.
\newblock Androzoo: Collecting millions of android apps for the research community.
\newblock In \emph{Proceedings of the 13th International Conference on Mining Software Repositories}, MSR '16, pages 468--471, New York, NY, USA, 2016. ACM.
\newblock ISBN 978-1-4503-4186-8.
\newblock \doi{10.1145/2901739.2903508}.
\newblock URL \url{http://doi.acm.org/10.1145/2901739.2903508}.

\bibitem[Alzaylaee et~al.(2020)Alzaylaee, Yerima, and Sezer]{alzaylaee2020dldroid}
Mohammed~K. Alzaylaee, Suleiman~Y. Yerima, and Sakir Sezer.
\newblock {DL-Droid: Deep learning based android malware detection using real devices}.
\newblock \emph{Comput. Secur.}, 89\penalty0 (C), feb 2020.
\newblock ISSN 0167-4048.
\newblock \doi{10.1016/j.cose.2019.101663}.
\newblock URL \url{https://doi-org.wwwproxy1.library.unsw.edu.au/10.1016/j.cose.2019.101663}.

\bibitem[Angelov(2020)]{angelov2020top2vec}
Dimo Angelov.
\newblock Top2vec: Distributed representations of topics.
\newblock \emph{arXiv preprint arXiv:2008.09470}, 2020.

\bibitem[{Apple}(2022)]{Apple}
{Apple}.
\newblock App security overview, May 2022.
\newblock URL \url{https://support.apple.com/en-au/guide/security/sec35dd877d0/web}.

\bibitem[Arp et~al.(2014)Arp, Spreitzenbarth, Hubner, Gascon, Rieck, and Siemens]{arp2014drebin}
Daniel Arp, Michael Spreitzenbarth, Malte Hubner, Hugo Gascon, Konrad Rieck, and CERT Siemens.
\newblock Drebin: Effective and explainable detection of android malware in your pocket.
\newblock In \emph{NDSS}, volume~14, pages 23--26, 2014.

\bibitem[Aslan and Samet(2020)]{malwareDetectionReview}
Ömer~Aslan Aslan and Refik Samet.
\newblock A comprehensive review on malware detection approaches.
\newblock \emph{IEEE Access}, 8:\penalty0 6249--6271, 2020.
\newblock \doi{10.1109/ACCESS.2019.2963724}.

\bibitem[Bianchi et~al.(2021)Bianchi, Terragni, and Hovy]{bianchi-etal-2021-pre}
Federico Bianchi, Silvia Terragni, and Dirk Hovy.
\newblock Pre-training is a hot topic: Contextualized document embeddings improve topic coherence.
\newblock In \emph{Proceedings of the 59th Annual Meeting of the Association for Computational Linguistics and the 11th International Joint Conference on Natural Language Processing (Volume 2: Short Papers)}, pages 759--766, Online, August 2021. Association for Computational Linguistics.
\newblock \doi{10.18653/v1/2021.acl-short.96}.
\newblock URL \url{https://aclanthology.org/2021.acl-short.96}.

\bibitem[Blei et~al.(2003)Blei, Ng, and Jordan]{lda2003}
David~M. Blei, Andrew~Y. Ng, and Michael~I. Jordan.
\newblock {Latent Dirichlet Allocation}.
\newblock \emph{J. Mach. Learn. Res.}, 3\penalty0 (null):\penalty0 993–1022, mar 2003.
\newblock ISSN 1532-4435.

\bibitem[Book et~al.(2013)Book, Pridgen, and Wallach]{book2013longitudinalanalysisandroidad}
Theodore Book, Adam Pridgen, and Dan~S. Wallach.
\newblock Longitudinal analysis of {A}ndroid ad library permissions, 2013.
\newblock URL \url{https://arxiv.org/abs/1303.0857}.

\bibitem[Calciati et~al.(2020)Calciati, Kuznetsov, Gorla, and Zeller]{permissions3}
Paolo Calciati, Konstantin Kuznetsov, Alessandra Gorla, and Andreas Zeller.
\newblock {Automatically Granted Permissions in Android apps: An Empirical Study on their Prevalence and on the Potential Threats for Privacy}.
\newblock In \emph{Proceedings of the 17th International Conference on Mining Software Repositories}, MSR '20, page 114–124, New York, NY, USA, 2020. Association for Computing Machinery.
\newblock ISBN 9781450375177.
\newblock \doi{10.1145/3379597.3387469}.
\newblock URL \url{https://doi-org.wwwproxy1.library.unsw.edu.au/10.1145/3379597.3387469}.

\bibitem[Church and Hanks(1990)]{church-hanks-1990-word}
Kenneth~Ward Church and Patrick Hanks.
\newblock Word association norms, mutual information, and lexicography.
\newblock \emph{Computational Linguistics}, 16\penalty0 (1):\penalty0 22--29, 1990.
\newblock URL \url{https://aclanthology.org/J90-1003}.

\bibitem[Delaney(2024)]{news24}
Ailish Delaney.
\newblock Delete now: The 17 dangerous apps blackmailing {A}ndroid, {A}pple phone users.
\newblock Technical report, https://apple.news/AhVoMPVr6Sq270cbnKfLzew, 2024.

\bibitem[Densons(2011)]{androguard}
Anthony Densons, Feb 2011.
\newblock URL \url{https://github.com/androguard/androguard}.

\bibitem[Devlin(2018)]{devlin2018bert}
Jacob Devlin.
\newblock {BERT}: Pre-training of deep bidirectional transformers for language understanding.
\newblock \emph{arXiv preprint arXiv:1810.04805}, 2018.

\bibitem[Enck et~al.(2014)Enck, Gilbert, Han, Tendulkar, Chun, Cox, Jung, McDaniel, and Sheth]{taintDroid}
William Enck, Peter Gilbert, Seungyeop Han, Vasant Tendulkar, Byung-Gon Chun, Landon~P. Cox, Jaeyeon Jung, Patrick McDaniel, and Anmol~N. Sheth.
\newblock Taintdroid: An information-flow tracking system for realtime privacy monitoring on smartphones.
\newblock \emph{ACM Trans. Comput. Syst.}, 32\penalty0 (2), jun 2014.
\newblock ISSN 0734-2071.
\newblock \doi{10.1145/2619091}.

\bibitem[Felt et~al.(2011)Felt, Chin, Hanna, Song, and Wagner]{demistyfiedAndroid}
Adrienne~Porter Felt, Erika Chin, Steve Hanna, Dawn Song, and David Wagner.
\newblock Android permissions demystified.
\newblock In \emph{Proceedings of the 18th ACM Conference on Computer and Communications Security}, CCS '11, page 627–638, New York, NY, USA, 2011. Association for Computing Machinery.
\newblock \doi{10.1145/2046707.2046779}.

\bibitem[Google()]{Google}
Google.
\newblock Google play protect, 2023.
\newblock URL \url{https://developers.google.com/android/play-protect}.

\bibitem[Gorla et~al.(2014)Gorla, Tavecchia, Gross, and Zeller]{gorla2014CHABADA}
Alessandra Gorla, Ilaria Tavecchia, Florian Gross, and Andreas Zeller.
\newblock Checking app behavior against app descriptions.
\newblock In \emph{Proceedings of the 36th International Conference on Software Engineering}, page 1025–1035. Association for Computing Machinery, 2014.
\newblock ISBN 9781450327565.
\newblock URL \url{https://doi.org/10.1145/2568225.2568276}.

\bibitem[Grootendorst(2022)]{grootendorst2022bertopic}
Maarten Grootendorst.
\newblock {BERTopic}: Neural topic modeling with a class-based {TF-IDF} procedure, 2022.

\bibitem[Karunanayake et~al.(2020)Karunanayake, Rajasegaran, Gunathillake, Seneviratne, and Jourjon]{karunanayake2020multi}
Naveen Karunanayake, Jathushan Rajasegaran, Ashanie Gunathillake, Suranga Seneviratne, and Guillaume Jourjon.
\newblock A multi-modal neural embeddings approach for detecting mobile counterfeit apps: A case study on {Google Play} store.
\newblock \emph{IEEE Transactions on Mobile Computing}, 21\penalty0 (1):\penalty0 16--30, 2020.

\bibitem[kaspersky23()]{Kaspersky}
kaspersky23.
\newblock Attacks on mobile devices significantly increase in 2023, 2023.
\newblock URL \url{https://www.kaspersky.com/about/press-releases/2024_attacks-on-mobile-devices-significantly-increase-in-2023}.

\bibitem[Lau et~al.(2010)Lau, Newman, Karimi, and Baldwin]{newman2010}
Jey~Han Lau, David Newman, Sarvnaz Karimi, and Timothy Baldwin.
\newblock Best topic word selection for topic labelling.
\newblock In \emph{Proceedings of the 23rd International Conference on Computational Linguistics: Posters}, COLING '10, page 605–613, USA, 2010. Association for Computational Linguistics.

\bibitem[Lau et~al.(2014)Lau, Newman, and Baldwin]{lau-etal-2014-machine}
Jey~Han Lau, David Newman, and Timothy Baldwin.
\newblock Machine reading tea leaves: Automatically evaluating topic coherence and topic model quality.
\newblock In Shuly Wintner, Sharon Goldwater, and Stefan Riezler, editors, \emph{Proceedings of the 14th Conference of the {E}uropean Chapter of the Association for Computational Linguistics}, pages 530--539, Gothenburg, Sweden, April 2014. Association for Computational Linguistics.
\newblock \doi{10.3115/v1/E14-1056}.
\newblock URL \url{https://aclanthology.org/E14-1056}.

\bibitem[Le and Mikolov(2014)]{le2014ddoc2vec}
Quoc Le and Tomas Mikolov.
\newblock Distributed representations of sentences and documents.
\newblock In \emph{International conference on machine learning}, pages 1188--1196. PMLR, 2014.

\bibitem[Mahdavifar et~al.(2020)Mahdavifar, Abdul~Kadir, Fatemi, Alhadidi, and Ghorbani]{maldroid2020}
Samaneh Mahdavifar, Andi~Fitriah Abdul~Kadir, Rasool Fatemi, Dima Alhadidi, and Ali~A. Ghorbani.
\newblock Dynamic {A}ndroid malware category classification using semi-supervised deep learning.
\newblock In \emph{2020 IEEE 18th Intl Conf on Dependable, Autonomic and Secure Computing, Intl Conf on Pervasive Intelligence and Computing, Intl Conf on Cloud and Big Data Computing, Intl Conf on Cyber Science and Technology Congress (DASC/PiCom/CBDCom/CyberSciTech)}, pages 515--522, 2020.
\newblock \doi{10.1109/DASC-PICom-CBDCom-CyberSciTech49142.2020.00094}.

\bibitem[Mariconti et~al.(2016)Mariconti, Onwuzurike, Andriotis, De~Cristofaro, Ross, and Stringhini]{mariconti2016mamadroid}
Enrico Mariconti, Lucky Onwuzurike, Panagiotis Andriotis, Emiliano De~Cristofaro, Gordon Ross, and Gianluca Stringhini.
\newblock Mamadroid: Detecting android malware by building {M}arkov chains of behavioral models.
\newblock \emph{arXiv preprint arXiv:1612.04433}, 2016.

\bibitem[McInnes et~al.(2018)McInnes, Healy, and Melville]{mcinnes2018umap}
Leland McInnes, John Healy, and James Melville.
\newblock {UMAP}: Uniform manifold approximation and projection for dimension reduction.
\newblock \emph{arXiv preprint arXiv:1802.03426}, 2018.

\bibitem[Meng et~al.(2022)Meng, Zhang, Huang, Zhang, and Han]{topclus2022}
Yu~Meng, Yunyi Zhang, Jiaxin Huang, Yu~Zhang, and Jiawei Han.
\newblock Topic discovery via latent space clustering of pretrained language model representations.
\newblock In \emph{Proceedings of the ACM Web Conference 2022}, WWW '22, page 3143–3152, New York, NY, USA, 2022. Association for Computing Machinery.
\newblock ISBN 9781450390965.
\newblock \doi{10.1145/3485447.3512034}.
\newblock URL \url{https://doi-org.wwwproxy1.library.unsw.edu.au/10.1145/3485447.3512034}.

\bibitem[Pan et~al.(2018)Pan, Ren, Lindorfer, Wilson, and Choffnes]{permissions2}
Elleen Pan, Jingjing Ren, Martina Lindorfer, Christo Wilson, and David Choffnes.
\newblock Panoptispy: Characterizing audio and video exfiltration from {A}ndroid applications.
\newblock \emph{Proceedings on Privacy Enhancing Technologies}, 2018.

\bibitem[Peddinti et~al.(2019)Peddinti, Bilogrevic, Taft, Pelikan, Erlingsson, Anthonysamy, and Hogben]{permissions1}
Sai~Teja Peddinti, Igor Bilogrevic, Nina Taft, Martin Pelikan, \'{U}lfar Erlingsson, Pauline Anthonysamy, and Giles Hogben.
\newblock Reducing permission requests in mobile apps.
\newblock In \emph{Proceedings of the Internet Measurement Conference}, IMC '19, page 259–266, New York, NY, USA, 2019. Association for Computing Machinery.
\newblock ISBN 9781450369480.
\newblock \doi{10.1145/3355369.3355584}.
\newblock URL \url{https://doi.org/10.1145/3355369.3355584}.

\bibitem[Rajasegaran et~al.(2019)Rajasegaran, Karunanayake, Gunathillake, Seneviratne, and Jourjon]{rajasegaran2019multi}
Jathushan Rajasegaran, Naveen Karunanayake, Ashanie Gunathillake, Suranga Seneviratne, and Guillaume Jourjon.
\newblock A multi-modal neural embeddings approach for detecting mobile counterfeit apps.
\newblock In \emph{The World Wide Web Conference}, pages 3165--3171, 2019.

\bibitem[R\"{o}der et~al.(2015)R\"{o}der, Both, and Hinneburg]{roder2015}
Michael R\"{o}der, Andreas Both, and Alexander Hinneburg.
\newblock Exploring the space of topic coherence measures.
\newblock In \emph{Proceedings of the Eighth ACM International Conference on Web Search and Data Mining}, WSDM '15, page 399–408, New York, NY, USA, 2015. Association for Computing Machinery.
\newblock ISBN 9781450333177.
\newblock \doi{10.1145/2684822.2685324}.
\newblock URL \url{https://doi.org/10.1145/2684822.2685324}.

\bibitem[Sayed et~al.(2023)Sayed, Bra{\c{s}}oveanu, Nixon, and Scharl]{bertopicNews}
Mohamad~Al Sayed, Adrian M.~P. Bra{\c{s}}oveanu, Lyndon J.~B. Nixon, and Arno Scharl.
\newblock Unsupervised topic modeling with {BERTopic }for coarse and fine-grained news classification.
\newblock In Ignacio Rojas, Gonzalo Joya, and Andreu Catala, editors, \emph{Advances in Computational Intelligence}, pages 162--174, Cham, 2023. Springer Nature Switzerland.
\newblock ISBN 978-3-031-43085-5.

\bibitem[Seneviratne et~al.(2015)Seneviratne, Seneviratne, Kaafar, Mahanti, and Mohapatra]{seneviratne2015early-detection}
Suranga Seneviratne, Aruna Seneviratne, Mohamed~Ali Kaafar, Anirban Mahanti, and Prasant Mohapatra.
\newblock Early detection of spam mobile apps.
\newblock In \emph{Proceedings of the 24th International Conference on World Wide Web}, pages 949--959, 2015.

\bibitem[Seneviratne et~al.(2017)Seneviratne, Seneviratne, Kaafar, Mahanti, and Mohapatra]{seneviratne2017spam}
Suranga Seneviratne, Aruna Seneviratne, Mohamed~Ali Kaafar, Anirban Mahanti, and Prasant Mohapatra.
\newblock Spam mobile apps: Characteristics, detection, and in the wild analysis.
\newblock \emph{ACM Transactions on the Web (TWEB)}, 11\penalty0 (1):\penalty0 1--29, 2017.

\bibitem[Shabtai et~al.(2012)Shabtai, Kanonov, Elovici, Glezer, and Weiss]{shabtai2012andromaly}
Asaf Shabtai, Uri Kanonov, Yuval Elovici, Chanan Glezer, and Yael Weiss.
\newblock {Andromaly}: {A} behavioral malware detection framework for {A}ndroid devices.
\newblock \emph{Journal of Intelligent Information Systems}, 38\penalty0 (1):\penalty0 161--190, 2012.

\bibitem[Surian et~al.(2017)Surian, Seneviratne, Seneviratne, and Chawla]{surian2017app-miscategorization}
Didi Surian, Suranga Seneviratne, Aruna Seneviratne, and Sanjay Chawla.
\newblock App miscategorization detection: A case study on {Google Play}.
\newblock \emph{IEEE Transactions on Knowledge and Data Engineering}, 29\penalty0 (8):\penalty0 1591--1604, 2017.

\bibitem[Toulas(2023)]{news23}
Bill Toulas.
\newblock Android malware infiltrates 60 {Google Play} apps with 100m installs.
\newblock Technical report, https://www.bleepingcomputer.com/news/security/android-malware-infiltrates-60-google-play-apps-with-100m-installs/, 2023.

\bibitem[Turan et~al.(2024)Turan, Yildiz, and B{\"u}y{\"u}ktanir]{turan2024}
Salih~Can Turan, Kaz{\i}m Yildiz, and B{\"u}{\c{s}}ra B{\"u}y{\"u}ktanir.
\newblock Comparison of {LDA}, {NMF} and {BERTopic} topic modeling techniques on amazon product review dataset: A case study.
\newblock In \emph{Computing, Internet of Things and Data Analytics}, pages 23--31. Springer Nature Switzerland, 2024.
\newblock ISBN 978-3-031-53717-2.

\bibitem[Wang et~al.(2023)Wang, Chen, Chen, and Chen]{wang2023identifying}
Zhongyi Wang, Jing Chen, Jiangping Chen, and Haihua Chen.
\newblock Identifying interdisciplinary topics and their evolution based on {BERTopic}.
\newblock \emph{Scientometrics}, pages 1--26, 2023.

\bibitem[Wu et~al.(2012)Wu, Mao, Wei, Lee, and Wu]{droidmat2012}
Dong-Jie Wu, Ching-Hao Mao, Te-En Wei, Hahn-Ming Lee, and Kuo-Ping Wu.
\newblock {DroidMat}: Android malware detection through manifest and {API} calls tracing.
\newblock In \emph{2012 Seventh Asia Joint Conference on Information Security}, pages 62--69, 2012.
\newblock \doi{10.1109/AsiaJCIS.2012.18}.

\bibitem[Wu et~al.(2024{\natexlab{a}})Wu, Nguyen, and Luu]{wu2024}
Xiaobao Wu, Thong Nguyen, and Anh~Tuan Luu.
\newblock A survey on neural topic models: methods, applications, and challenges.
\newblock \emph{Artificial Intelligence Review}, 57\penalty0 (2):\penalty0 18, 2024{\natexlab{a}}.

\bibitem[Wu et~al.(2024{\natexlab{b}})Wu, Nguyen, Zhang, Wang, and Luu]{wu2024fastopicfastadaptivestable}
Xiaobao Wu, Thong Nguyen, Delvin~Ce Zhang, William~Yang Wang, and Anh~Tuan Luu.
\newblock {FASTopic:} a fast, adaptive, stable, and transferable topic modeling paradigm, 2024{\natexlab{b}}.
\newblock URL \url{https://arxiv.org/abs/2405.17978}.

\bibitem[Yan and Yin(2012)]{droidscope2012}
Lok~Kwong Yan and Heng Yin.
\newblock {DroidScope:} {S}eamlessly reconstructing the os and {Dalvik} semantic views for dynamic {A}ndroid malware analysis.
\newblock In \emph{Proceedings of the 21st USENIX Conference on Security Symposium}, Security'12, page~29, USA, 2012. USENIX Association.

\bibitem[Yuan et~al.(2014)Yuan, Lu, Wang, and Xue]{droidsec2014}
Zhenlong Yuan, Yongqiang Lu, Zhaoguo Wang, and Yibo Xue.
\newblock {Droid-Sec:} {D}eep learning in {A}ndroid malware detection.
\newblock \emph{SIGCOMM Comput. Commun. Rev.}, 44\penalty0 (4):\penalty0 371–372, aug 2014.
\newblock ISSN 0146-4833.
\newblock \doi{10.1145/2740070.2631434}.
\newblock URL \url{https://doi-org.wwwproxy1.library.unsw.edu.au/10.1145/2740070.2631434}.

\end{thebibliography}

\appendix

\section{Qualitative Analysis of Topic Assignments}\label{appendix:topic_wordclouds}

To validate the effectiveness of BERTopic in assigning a single dominant topic to app descriptions, we present qualitative examples of topic assignments for two popular messaging apps, Viber and Telegram. Figures~\ref{fig:bertopic_49_topic}, \ref{fig:telegram_lda_topics}, and \ref{fig:chabada_topics} show word clouds for topics assigned using BERTopic, Latent Dirichlet Allocation (LDA), and CHABADA, respectively.

As seen in Figure~\ref{fig:bertopic_49_topic}, BERTopic identifies a coherent and interpretable primary topic capturing core app features, with terms like “message,” “chat,” “call,” and “friend,” reflecting the messaging and communication functionalities of Viber and Telegram. In contrast, LDA and CHABADA assign multiple topics per app, resulting in fragmented representations where some topics lack clear relevance.

\begin{figure}[htbp]
\centering
\includegraphics[scale=0.15]{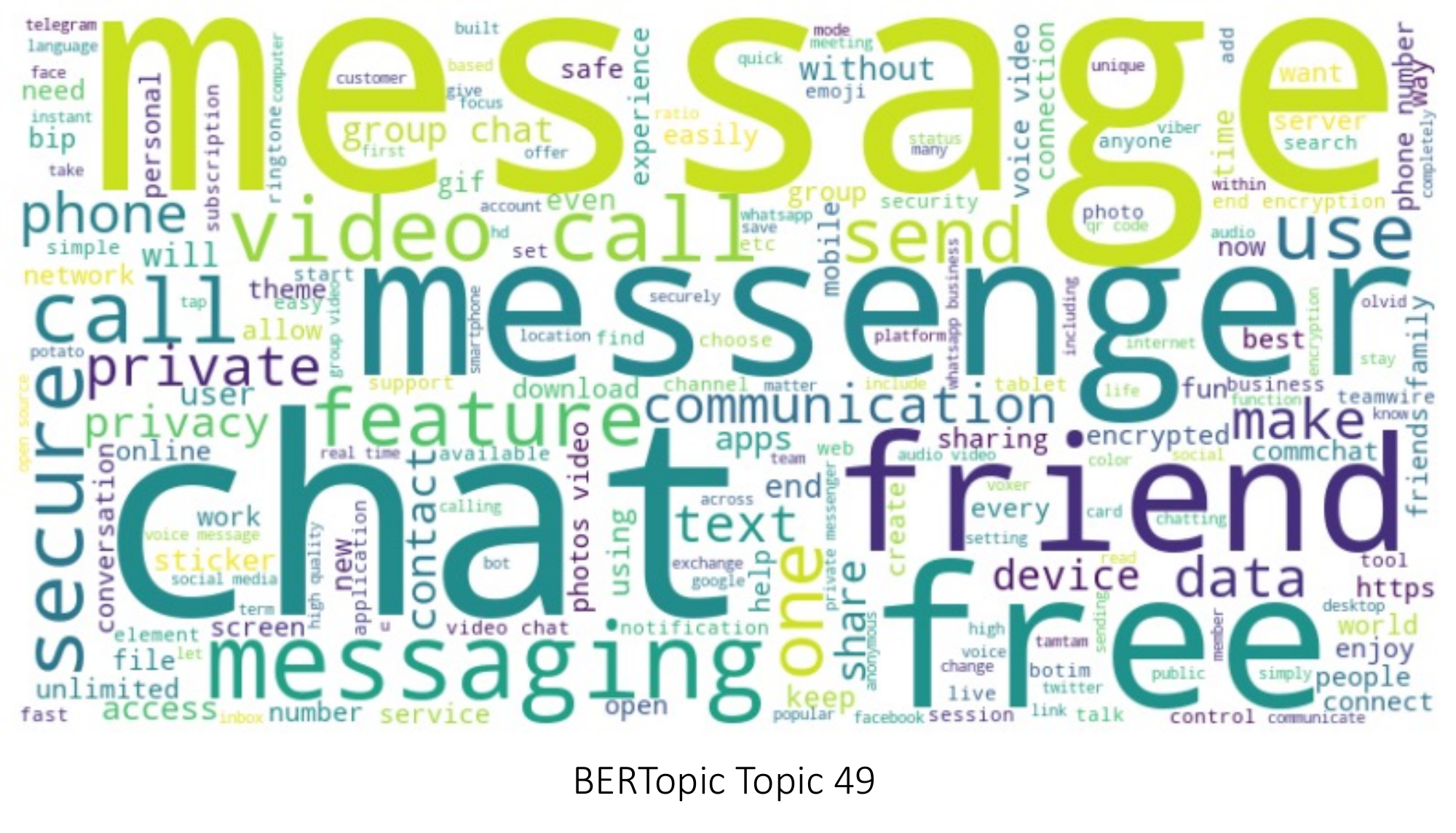}
\vspace{-2mm}
\caption{For both Viber and Telegram apps, BERTopic assigns a dominant topic with terms closely related to messaging. The high frequency of relevant terms indicates that BERTopic successfully identifies and isolates the core features of messaging apps, making the assignment intuitive and directly interpretable.}
\Description{}
\label{fig:bertopic_49_topic}
\end{figure}

For example, as shown in Figure~\ref{fig:telegram_lda_topics}, LDA assigns multiple topics to Telegram (e.g., Topics 24, 23, 9, and 39). Among these, only one topic closely relates to messaging, while others include generic terms like “device”, “secure”, and “time”, diluting interpretability. 
CHABADA’s approach, which applies k-means clustering to LDA topics, produces broad clusters that lack focus, as shown in Figure~\ref{fig:chabada_topics}. These clusters often include generic terms like “make,” “help,” “one,” and “use,” making it difficult to identify specific functionalities.

These examples highlight the advantage of our approach, which selects the single most meaningful topic from BERTopic as a reliable representation for each app. By focusing on one coherent topic, we enhance interpretability and reduce the complexity and ambiguity associated with multi-topic assignments.

\begin{figure}[htbp]
\centering
\includegraphics[scale=0.15]{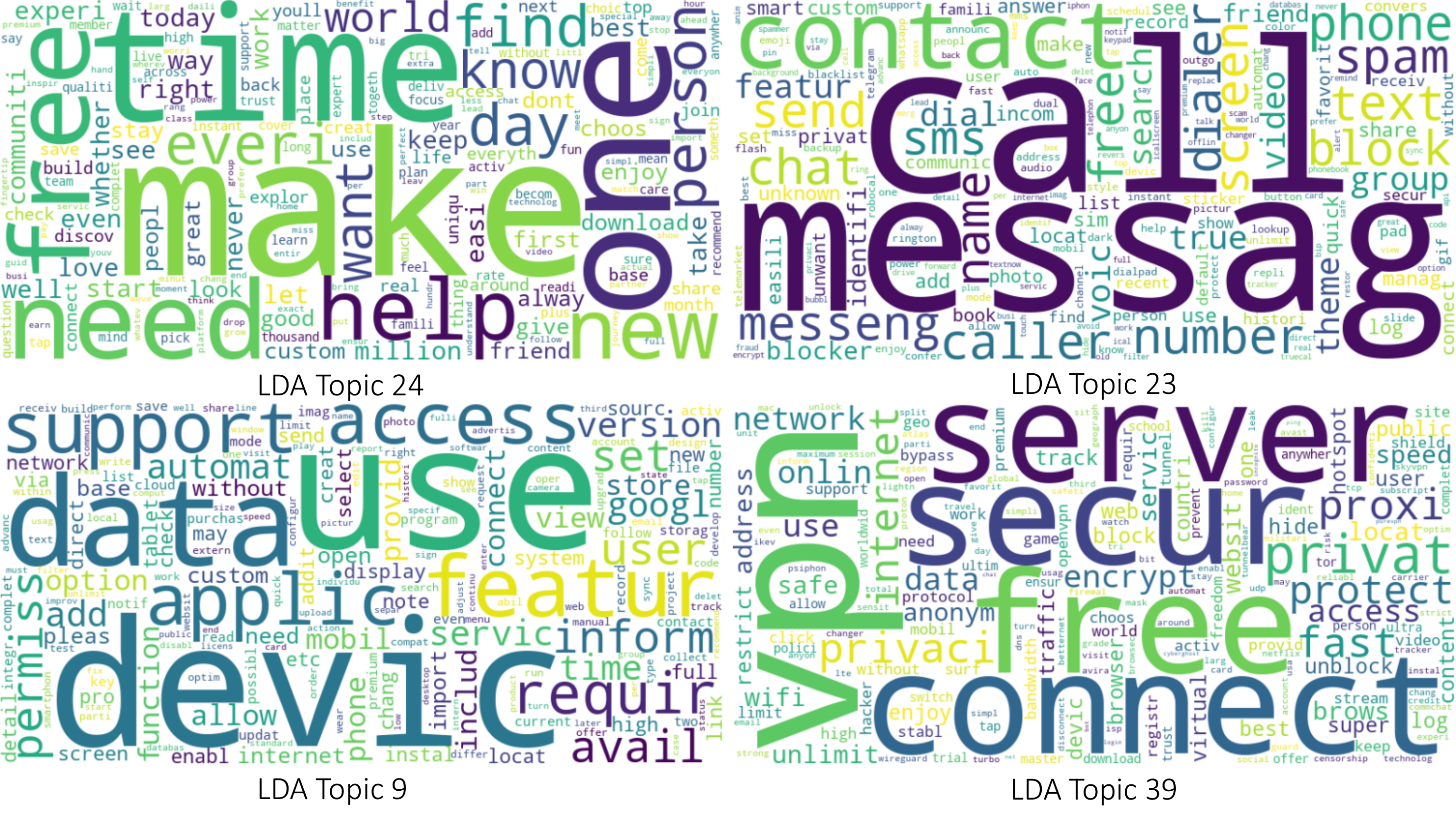}
\vspace{-2mm}
\caption{Wordclouds of topics assigned by LDA to Telegram. Here, while Topic 23 is more relevant to Telegram, the applicability of other topics; 9, 24, and 39 is marginal. They are still related to the app. However, they don't show a strong coherence in what each topic is representing.}
\Description{}
\label{fig:telegram_lda_topics}
\end{figure}

\begin{figure}[htbp]
\centering
\includegraphics[scale=0.15]{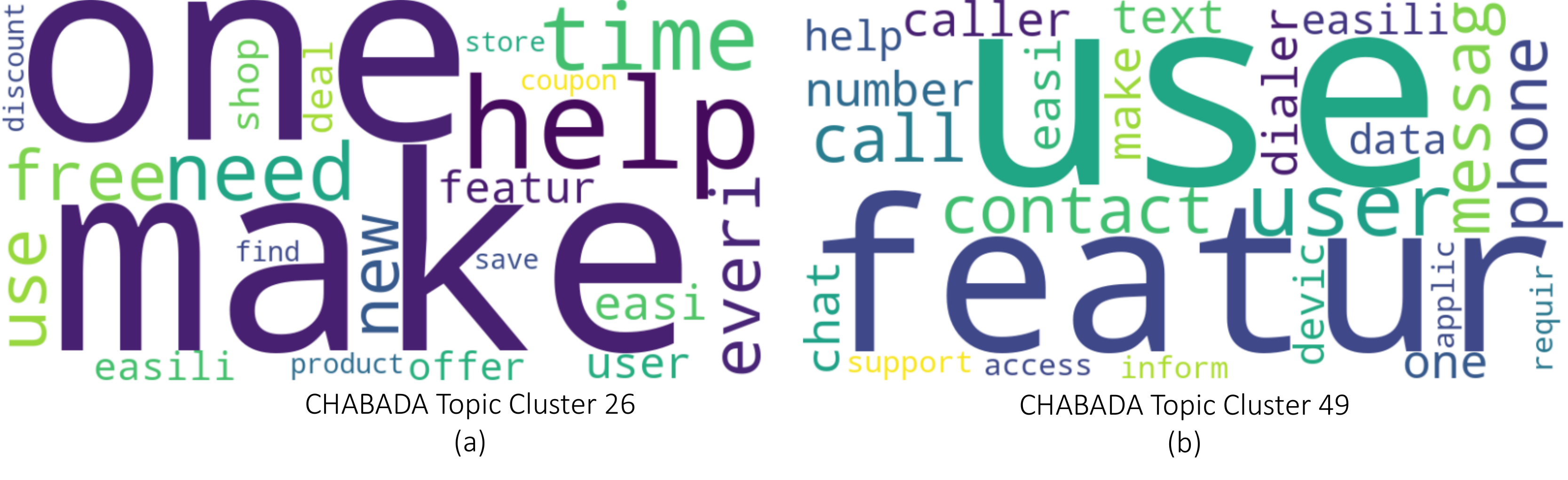}
\vspace{-2mm}
\caption{(a) Word cloud of the topic cluster assigned by CHABADA to Telegram’s app description, and (b) word cloud of the topic cluster assigned by CHABADA to Viber’s app description. Both clusters predominantly consist of generic terms that fail to capture the distinct purposes of the apps. While some keywords may loosely align with the app’s features, the clusters remain too broad to effectively represent the unique messaging and communication functionalities.}
\Description{}
\label{fig:chabada_topics}
\end{figure}

\section{Topic Coherence Metrics}\label{appendix:metrics}

{\bf \textit{Normalized Pointwise Mutual Information (NPMI)}} is a robust coherence measure for topic modeling, addressing the low-frequency bias of Pointwise Mutual Information (PMI) by introducing a fixed upper bound. PMI measures how often two terms $w_i$ and $w_j$ co-occur compared to their independent occurrences~\cite{church-hanks-1990-word}. \textit{NPMI} normalizes PMI by dividing it by the negative logarithm of the co-occurrence probability, ensuring more reliable coherence scores that reflect semantic relationships~\cite{lau-etal-2014-machine} (see Equation 1). \textit{NPMI} ranges from -1 to 1, with higher values indicating better coherence.

\begin{equation}\label{NPMI_calc}
    \text{NPMI}(w_i) = \sum_{j=1}^{N-1} \frac{\log \frac{P(w_i, w_j)}{P(w_i) P(w_j)}}{-\log P(w_i, w_j)}
\end{equation}

{\bf \textit{Cv score}} is a widely used coherence metric, showing the highest correlation with human evaluations according to Röder et al.~\cite{roder2015}. To compute the \textit{Cv} score, the corpus is tokenized, and top-N words are extracted for each topic. Word embeddings (e.g., Word2Vec or GloVe) represent these words in a high-dimensional space. A sliding window generates context vectors based on word co-occurrence, and cosine similarity is calculated between the context vectors of word pairs. These similarity scores are aggregated and normalized to produce the final \textit{Cv} score, which ranges from 0 to 1, with higher values indicating better topic coherence and stronger semantic relationships.

\end{document}